\documentclass[twocolumn,superscriptaddress]{revtex4-1}
\usepackage[pdftex]{graphicx}
\usepackage{siunitx}
\begin{document}
\title{Lattice constants and magnetism of $L1_0$-ordered FePt under high pressure}  
\author{S. Sawada}  
\affiliation{Department of Material Science, Graduate School of Science, University of Hyogo, Ako-gun, 
Hyogo 678-1297, Japan}
\author{K. Okai} 
\affiliation{Department of Material Science, Graduate School of Science, University of Hyogo, Ako-gun, 
Hyogo 678-1297, Japan}
\author{H. Fukui}
\affiliation{Department of Material Science, Graduate School of Science, University of Hyogo, Ako-gun, 
Hyogo 678-1297, Japan}
\affiliation{Japan Synchrotron Radiation Research Institute, Sayo, Hyogo 679-5198, Japan.}
\author{R. Takahashi}
\affiliation{Department of Material Science, Graduate School of Science, University of Hyogo, Ako-gun, 
Hyogo 678-1297, Japan}
\author{N. Ishimatsu}
\affiliation{Graduate School of Advanced Science and Engineering, Hiroshima University, Higashihiroshima, Hiroshima 739-8526, Japan}
\author{H. Maruyama}
\affiliation{Graduate School of Advanced Science and Engineering, Hiroshima University, Higashihiroshima, Hiroshima 739-8526, Japan}
\author{N.~Kawamura}
\affiliation{Japan Synchrotron Radiation Research Institute, 
Sayo, Hyogo 679-5198, Japan.}
\author{S.~Kawaguchi}
\affiliation{Japan Synchrotron Radiation Research Institute, 
Sayo, Hyogo 679-5198, Japan.}
\author{N. Hirao}
\affiliation{Japan Synchrotron Radiation Research Institute, 
Sayo, Hyogo 679-5198, Japan.}
\author{T. Seki}
\affiliation{Institute for Materials Research, Tohoku University, Sendai, Miyagi 980-8577, Japan}
\affiliation{Center for Spintronics Research Network, Tohoku University, Sendai, Miyagi 980-8577, Japan}
\author{K. Takanashi}
\affiliation{Institute for Materials Research, Tohoku University, Sendai, Miyagi 980-8577, Japan}
\affiliation{Center for Spintronics Research Network, Tohoku University, Sendai, Miyagi 980-8577, Japan}
\affiliation{Center for Science and Innovation in Spintronics, Tohoku University, Sendai, Miyagi 980-8577, Japan}
\author{S. Ohmura}
\affiliation{Faculty of Engineering, Hiroshima Institute of Technology, Hiroshima, 
Hiroshima 731-5193, Japan}
\author{H. Wadati}\email{wadati@sci.u-hyogo.ac.jp} 
\affiliation{Department of Material Science, Graduate School of Science, University of Hyogo, Ako-gun, 
Hyogo 678-1297, Japan}
\affiliation{Institute of Laser Engineering, Osaka University, Suita, Osaka 565-0871, Japan}
\begin{abstract}
We studied 
the relationship between the lattice constant and 
magnetism of $L1_0$-ordered FePt under high pressure 
by means of first-principles calculations and 
synchrotron x-ray measurements. 
Based on our calculations, we found that 
the $c/a$ ratio shows an anomaly at $\sim$ 20 GPa and 
that the Pt magnetic moment is sharply suppressed 
at $\sim$ 60 GPa. 
As for the $c/a$, we experimentally verified 
the anomaly at $\sim$ 20 GPa by powder x-ray diffraction. 
We also measured the x-ray magnetic circular dichroism 
at the Pt $L$ edge up to $\sim$ 20 GPa. 
Any significant change of the Pt magnetic moment was not observed, in agreement with the calculations. 
These results thus indicate the possibility 
that novel magnetic states can be created in $L1_0$-ordered FePt by lattice deformation under high pressure. 
\end{abstract}
\maketitle
\section{Introduction}
With the advent of a Super Smart Society, 
the amount of information which we need to handle 
in our daily life is rapidly increasing. 
For the storage of large amounts of data, 
there has been a lot of interest 
in recording technology with magnetic materials. 
In order to achieve an efficient way of 
magnetic recording, ferromagnets with a strong magneto-crystalline anisotropy (MCA) are highly desired. 
$L1_0$-ordered FePt has been intensively investigated 
due to its excellent chemical stability and 
strong uniaxial magnetic anisotropy ($K_u$) 
up to $\sim 7 \times 10^6$ J/$\mbox{m}^3$ \cite{oba,shima1,shima2,b2}. It is thus expected that this material will act as 
the next-generation magnetic recording devices. 

At ambient pressure, 
the magnetism of FePt has been extensively studied 
by first-principles calculations and experiments 
\cite{b4, sakuma, a3, b3}. 
From the first-principles calculations, 
it was found that the strong MCA in $L1_0$ FePt 
is due to the strong spin-orbit coupling (SOC) 
in the Pt $5d$ orbitals, and that the Pt atoms 
show ferromagnetism by the hybridization 
between the Fe $3d$ and Pt $5d$ states \cite{b4}. 
Ikeda {\it et al.} performed x-ray
magnetic circular dichroism (XMCD) measurements of
$L1_0$-ordered FePt thin films and determined 
the magnetic moments of Fe $3d$ and Pt $5d$ states \cite{b3}.

However, it is unknown how the lattice constants 
and magnetism change in $L1_0$ FePt under high pressure. 
Ko {\it et al.} determined the change of the unit-cell volume up to $\sim$ 55 GPa by performing 
synchrotron radiation powder X-ray diffraction (XRD) 
measurements of FePt alloys. 
They fitted the third-order Birch Murnaghan equation of 
state (BM-EOS) \cite{b13,b14} to the experimental data 
and obtained 
the parameters of $K_0=264.1$ GPa, $K^{\prime}=5.0$, 
and $V_0 = 26.74$ $\mbox{\AA}^3$ \cite{b7}, 
where 
$K_0$ is a reference bulk modulus, 
$K^{\prime}$ is its pressure derivative, and 
$V_0$ is a reference volume. 
In their work, the behaviors of the lattice constants 
$a$ and $c$ were not given and 
the degree of the $L1_0$ ordering parameter 
of the studied sample was not clear. 

In the present work, we studied the pressure effect 
of the lattice constants and magnetism 
of $L1_0$-ordered FePt 
through both first-principles calculations 
and synchrotron x-ray measurements. 
In the calculations, we simulated 
the lattice constants $a$ and $c$ and unit-cell volume 
as a function of pressure. 
Specifically, the $c/a$ ratio shows an anomaly 
at $\sim$ 20 GPa. Moreover, 
the Pt magnetic moment is sharply decreased 
at $\sim$ 60 GPa. 
In order to experimentally verify these anomalous 
behaviors indicated by the calculations, 
we performed measurements of 
powder XRD and XMCD under high pressure. 
We observed the $c/a$ anomaly at $\sim$ 20 GPa 
and almost no pressure dependence 
of the XMCD intensity up to $\sim$ 20 GPa. 
These results will open a door to controlling 
magnetic states in $L1_0$-ordered FePt 
by applying lattice deformation under high pressure. 

\section{Calculation}
We performed first-principles calculations 
by using the Quantum ESPRESSO package \cite{b18, b19} 
with a projector augmented wave (PAW) method \cite{b17} 
within the generalized gradient approximation (GGA) 
of the density functional theory. 
We used a $k$-point sampling of 
$24 \times 24 \times 24$ with Monkhorst-Pack 
integration scheme \cite{b20} and 
energy cutoff of 280 Ry for one 
self-consistent field cycle. 

Figure 1 shows the $L1_0$ crystal structure of FePt, where 
the solid lines represent the fcc-like unit cell and 
the dashed lines represent the primitive cell 
used for first-principles calculations. 
The $L1_0$ structure consists of alternating layers 
of Fe and Pt atoms along the (001) directions. 
The lattice parameters $a_t$ and $c_t$ for calculations 
are related to the fcc-like $a$ and $c$ 
as $a_t = a/\sqrt{2}$ and $c_t=c$. 
With fixing the unit cell volume, 
we searched for the minimum energy 
by varying $a_t$ and $c_t$. 
The pressure is 
estimated based on the mean of the three 
diagonal components of the stress tensor.

\begin{figure}
\centering
\includegraphics[width=70mm]{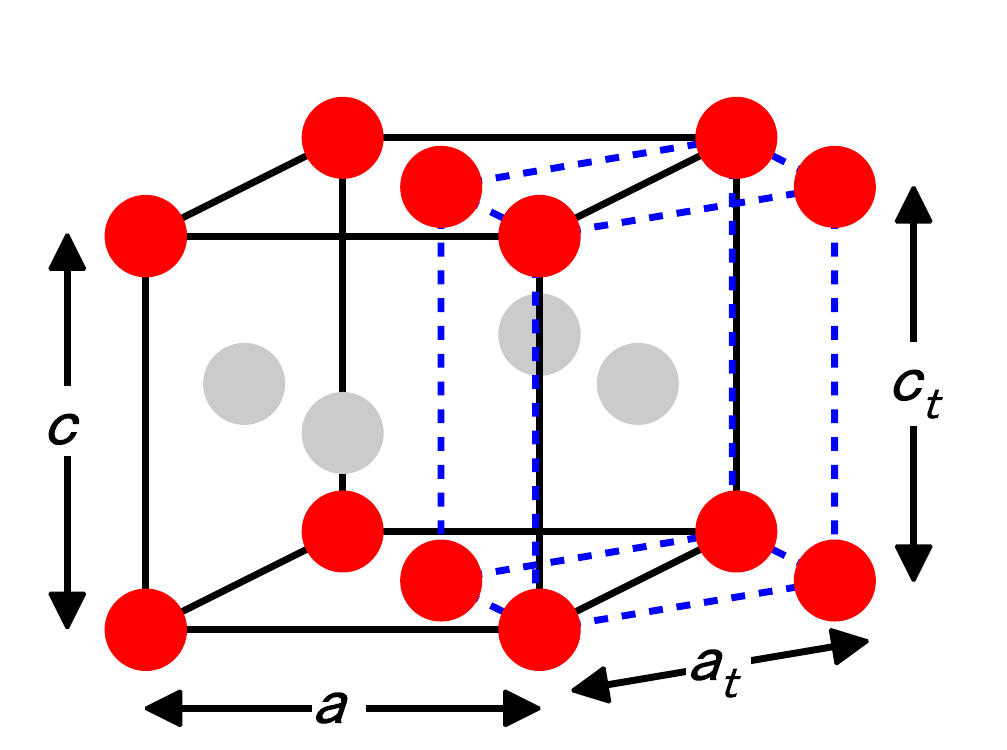}
\caption{The $L1_0$ crystal structure of FePt. 
The red (dark) and gray (light) spheres 
represent Fe and Pt atoms, respectively. 
The solid lines represent the fcc-like unit cell. 
The dashed lines represent the primitive cell 
used for first-principles calculations.}	
\label{fig1}
\end{figure}

We took spin polarization into account, and 
performed calculations 
in the presence and absence of SOC to investigate the 
effects of SOC on lattice constants and magnetism. 
In the absence of SOC, the valence electrons were  
Fe: $3s^23p^64s^23d^6$ and 
Pt: $5s^25p^65d^96s^1$, 
and the $c/a$ ratio step was set to 0.0005.
In the presence of SOC, the valence electrons were 
Fe: $3s^23p^64s^23d^6$ and 
Pt: $5s^25p^64f^{14}5d^96s^1$, 
and the $c/a$ ratio step was set to 0.0001. 

\section{Experiment}
The FePt bulk samples for XRD measurements 
were sintered at 1000 $^{\circ}$C for 24 hours. 
Then the sample was cut into a plate shape 
with $5 \times 5 \times 0.5$ $\mbox{mm}^3$ 
through the wet processing. 
The crystal structure was characterized 
with laboratory XRD (Cu $K\alpha$ line). 

We performed XRD and XMCD measurements 
at the SPring-8 beamlines BL10XU \cite{b15} 
and BL39XU, respectively. 
All the measurements were carried out 
at room temperature. 

The experimental conditions of XRD were as 
follows: x-ray wavelength of 0.4133 $\mbox{\AA}$ 
(the corresponding photon energy is 30 keV), 
no oscillation, and the exposure time of 5 seconds. 
We used a diamond-anvil cell with metallic 
rhenium gaskets to apply pressure to the sample. 
The culet size of the diamonds was 0.2 mm. 
In the sample chamber, we placed KCl, FePt, 
and KCl in this order. 
Powder XRD patterns were collected 
with the RIGAKU R-AXIS $\mbox{IV}^{++}$ image-plate detector. 
The pressure was quantitatively estimated based 
on the pressure-volume relationship of KCl 
reported in Ref.~\cite{b16}. 

The XMCD signals were collected by measuring 
the spectra of X-ray absorption spectroscopy (XAS) 
with the helicity-reversal method. 
A diamond phase retarder was employed 
to produce the circularly polarized beam. 
Powder FePt was annealed at 950 $^{\circ}$C for 84 hours 
for the XMCD measurements. 
The obtained powder sample was sieved with a 25 $\mu$m mesh, 
and the fine powder was loaded into a hole of the gasket 
by using methanol:ethanol ($=$ 4:1 ratio) solutions. 
The culet size of the diamonds was 0.45 mm, and nonmagnetic stainless steel (SU316L) gaskets 
were used to apply pressure to the sample.  
In order to magnetically saturate the sample, 
a high magnetic field 
of $\pm 10$ T was applied to the $L1_0$-FePt powder. 
We designed a miniature diamond anvil cell to be accommodated 
in the small sample chamber of the superconducting magnet 
(25 mm in diameter).  The pressure in the sample cavity was 
monitored by using the conventional method of ruby 
fluorescence measurements \cite{Ruby}.

\section{Results and discussions}
Figure 2 shows the pressure dependence of the 
lattice constants ($a$ and $c$), the $c/a$ ratio, 
and the primitive-cell volume ($V=a^2c/2$) 
obtained from the first-principles calculations, 
where the results both with and without SOC are shown. 
The lattice constants $a$ and $c$ decreased 
monotonically by the application of pressure, whereas 
the $c/a$ ratio showed non-monotonic behaviors, 
namely a local maximum 
at $\sim$ 10 GPa with SOC and 
at $\sim$ 20 GPa without SOC. 
We fitted the following BM-EOS to the obtained data
\begin{equation}
P(V)=\frac{3K_0}{2}(\eta^{-7}-\eta^{-5})
[1+\frac{3}{4}(K^{\prime}-4)(\eta^{-2}-1)],
\end{equation}
where $\eta =({V(P)/V_0})^{1/3}$, $({a(P)/a_0})$, 
or $({c(P)/c_0})$, 
and $P$ is the pressure. 
The subscript 0 means the value at the ambient conditions.  
The obtained parameters were 
$K_0=$ 210(3) GPa, $K^{\prime}=$ 4.5(1), 
and $V_0=$ 27.83(2) $\mbox{\AA}^3$ with SOC, and 
$K_0=$ 210(4) GPa, $K^{\prime}=$ 4.5(2), 
and $V_0=$ 28.04(2) $\mbox{\AA}^3$ without SOC. 

\begin{figure*}
\centering
\includegraphics[width=\linewidth]{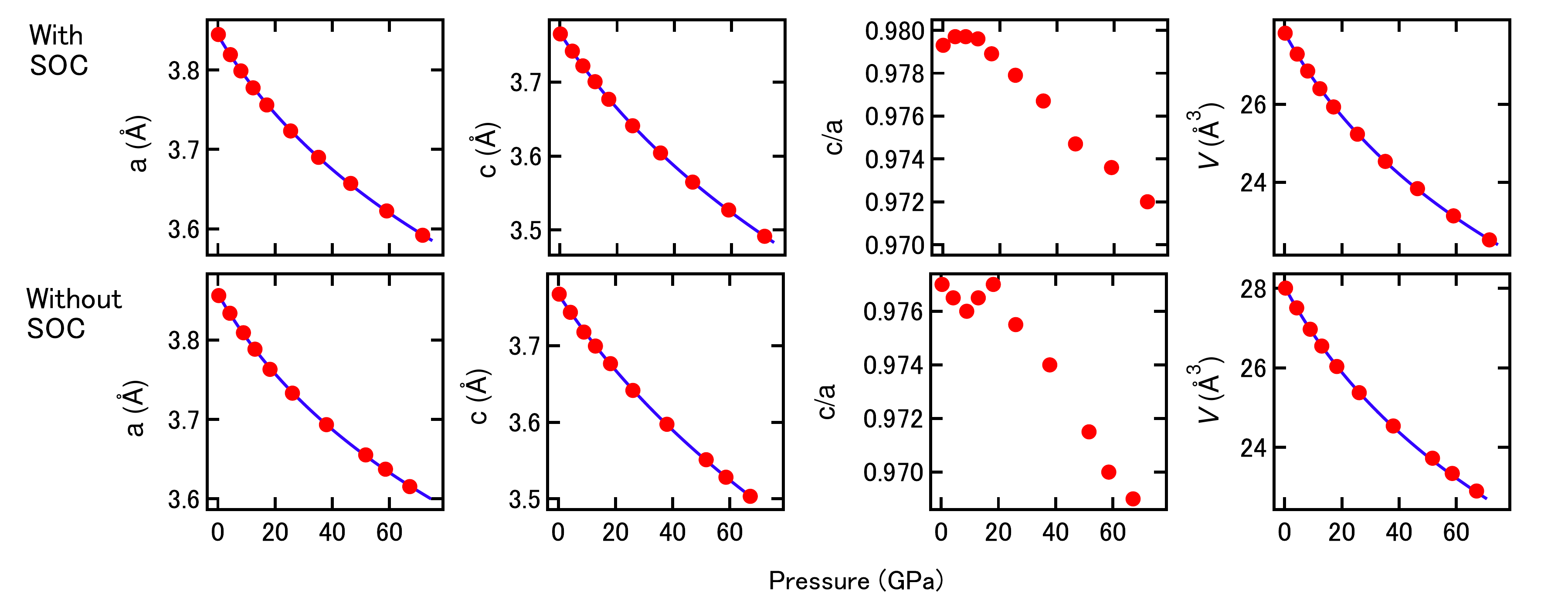}
\caption{Pressure dependence of lattice constants 
($a$ and $c$), the $c/a$ ratio, 
and the primitive-cell volume ($V$) 
estimated by the first-principles calculations. 
The solid lines are the fitted curves of 
the third-order BM-EOS. 
The results with and without SOC are shown.}
\label{fig2}
\end{figure*}

In order to investigate how the $c/a$ anomaly 
seen in Fig.~2 affects magnetism, 
we plotted the magnetic moments 
of Fe and Pt as a function of pressure in Fig.~3. 
At ambient pressure, the magnetic moments of Fe and Pt 
were estimated to be 3.0 $\mu_B$ and 0.37 $\mu_B$ 
with SOC, in good agreement with the experimental values 
reported in a prior XMCD study \cite{b3}. 
The Fe magnetic moment decreased monotonically 
with increasing pressure both with and without SOC. 
The slope of the decrease changes at $\sim$ 20 GPa, 
where we found that the $c/a$ ratio shows 
a local maximum from the calculations without SOC. 
The Pt magnetic moment was almost constant 
up to about 50 GPa with SOC and showed 
a local maximum at $\sim$ 10 GPa without SOC. 
Despite the difference between the two cases 
seen in the low-pressure regime, 
the magnetic moment suddenly diminished 
above 50 GPa in both cases. 

\begin{figure}
\centering
\includegraphics[width=\linewidth]{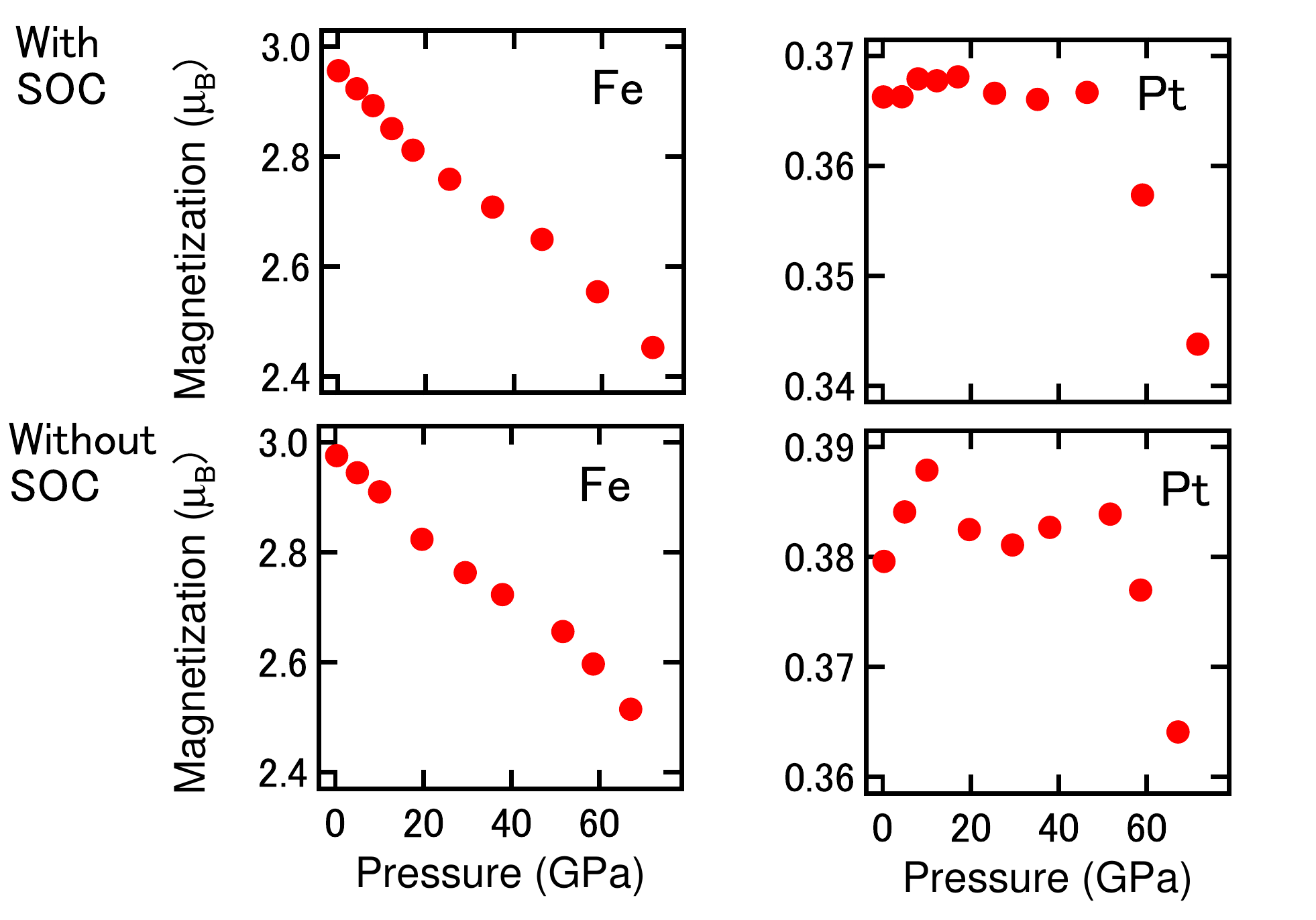}
\caption{Magnetic moments of Fe and Pt 
obtained by first-principles calculations 
under high pressure. 
The results with and without SOC are shown.}
\label{fig3}
\end{figure}

In the light of the pressure effects obtained 
by the first-principle calculations above, 
we aim to verify the obtained 
calculation results by performing x-ray measurements. 
First, we attempted to observe the c/a behavior 
by powder XRD. Before applying high pressure, 
we characterized the FePt sample 
by using laboratory XRD. 
Figure 4 (a) shows a powder XRD pattern 
at ambient pressure. 
We observed the peaks 
corresponding to the following 
Miller indices: 001, 110, 111, 200, 002, 201, 112, 
220, 202, 311, 222, and 312, mirroring 
the $L1_0$-ordered structure. 
Based on this XRD pattern, 
the lattice constants were determined 
to be $a=3.857$ $\mbox{\AA}$ and $c=3.735$ $\mbox{\AA}$, 
in good agreement with $a_t=2.722$ $\mbox{\AA}$ 
and $c_t=3.700$ $\mbox{\AA}$ reported in Ref.~\cite{yuasa}. 
The degree of long-range chemical order $S$ is defined 
as follows:
\begin{equation}
S=\frac{r_{Fe}-x_{Fe}}{y_{Fe}}=
\frac{r_{Pt}-x_{Pt}}{y_{Pt}},
\end{equation}
where $x_{\mathrm{el}}$, $y_{\mathrm{el}}$, 
and $r_{\mathrm{el}}$ (${\mathrm{el}}=$ Fe, Pt) 
are atomic fraction, fraction of the sites, 
and fraction of the sites occupied by the correct 
element, respectively. 
Based on the Rietveld analysis of the powder XRD pattern, 
we determined the value of $S$ to be 0.67. 

\begin{figure}
\centering
\includegraphics[width=\linewidth]{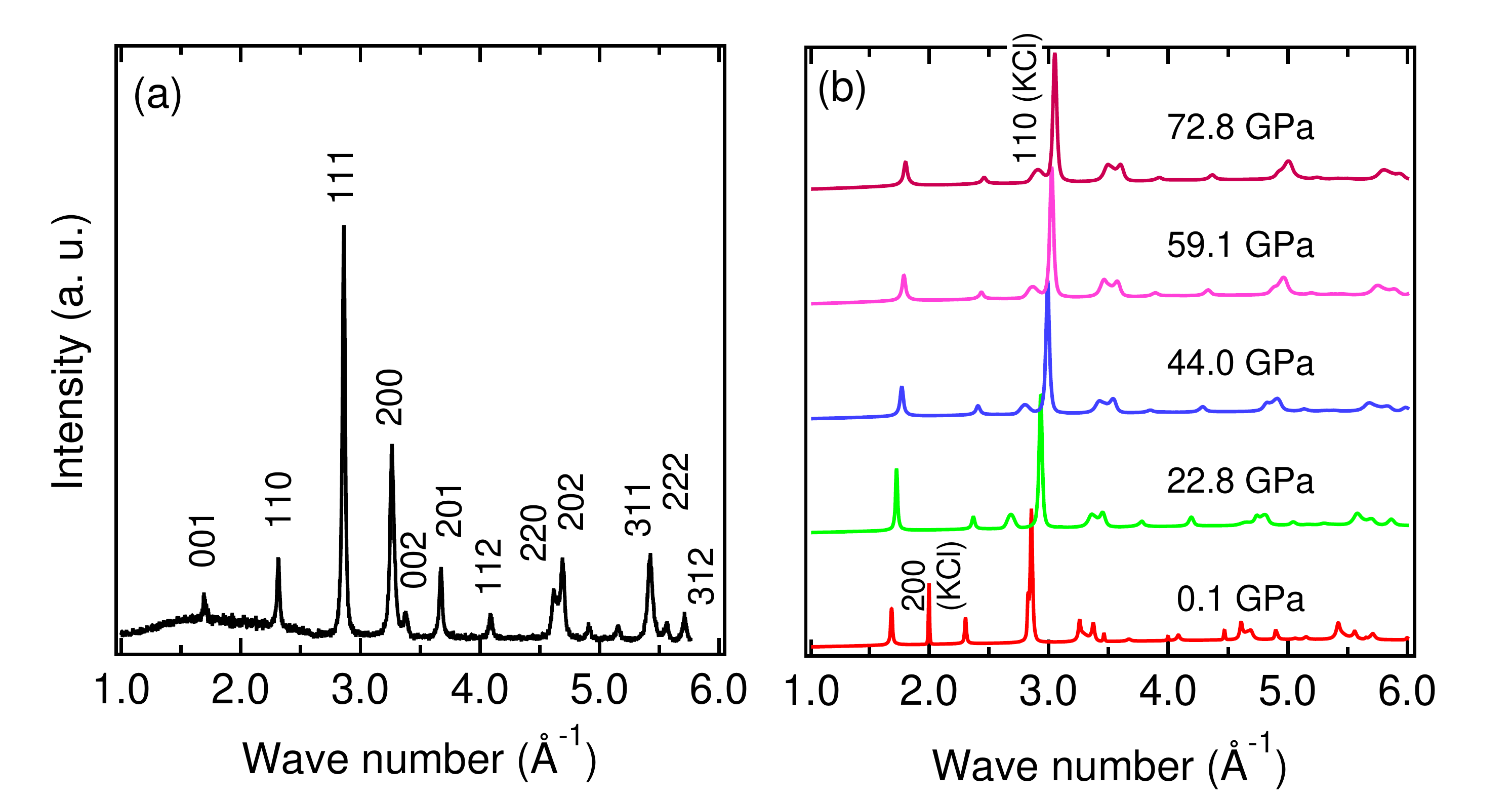}
\caption{Powder XRD patterns of FePt under ambient (a) 
and high pressure (b). In panel (b), 
the peaks of KCl were also observed. 
Indices in panel (b) are for KCl B1 (below 2.8 GPa) 
and B2 (above that) phases.}
\label{fig4}
\end{figure}

After these characterizations, we performed powder 
XRD measurements under high pressure. 
Figure 4 (b) shows the typical XRD patterns 
in the applications of pressure. 
The 200 and 110 peaks of KCl were observed 
at 0.1 GPa and at 22.8 GPa or more, respectively. 
This corresponds to the B1 to B2 phase transition 
of KCl at 2.6 GPa \cite{b16}. 
The pressure-independent XRD patterns of FePt indicate 
that the $L1_0$-ordered crystal structure with $S=0.67$ 
was kept up to 72.8 GPa. 
We converted XRD images into conventional XRD patterns 
by using IPAnalyzer and then determined the lattice constants 
$a$ and $c$ by using PDIndexer \cite{software}. 
The obtained parameters $a$, $c$, $c/a$ and $V$ 
are listed in TABLE I, and plotted in Fig.~5. 
The lattice constants $a$ and $c$ monotonically 
decreased by applying pressure. 
We fitted BM-EOS to the experimental data and 
obtained the parameters 
$K_0=$ 231(7) GPa, $K^{\prime}=$ 4.5(3), 
and $V_0=$ 27.64(3) $\mbox{\AA}^3$. 
Since there is no anomaly in $V$ as a function 
of pressure, we used a single parameter set. 
The obtained Birch-Murnaghan parameters 
are listed in Table II, 
alongside the values obtained from our
first-principles calculations above. 
The experimental value of $V_0$ is 
smaller than those of the calculations 
due to the tendency of GGA to systematically 
overestimate lattice constants \cite{bond}. 
The experimental $K$ is larger 
than the calculated values 
because GGA underestimates bulk moduli \cite{bond}. 
The experimental value of $c/a$ shows a maximum at 20 GPa, 
which is qualitatively in agreement with 
the calculation without SOC in Fig.~2. 

\begin{figure}
\centering
\includegraphics[width=\linewidth]{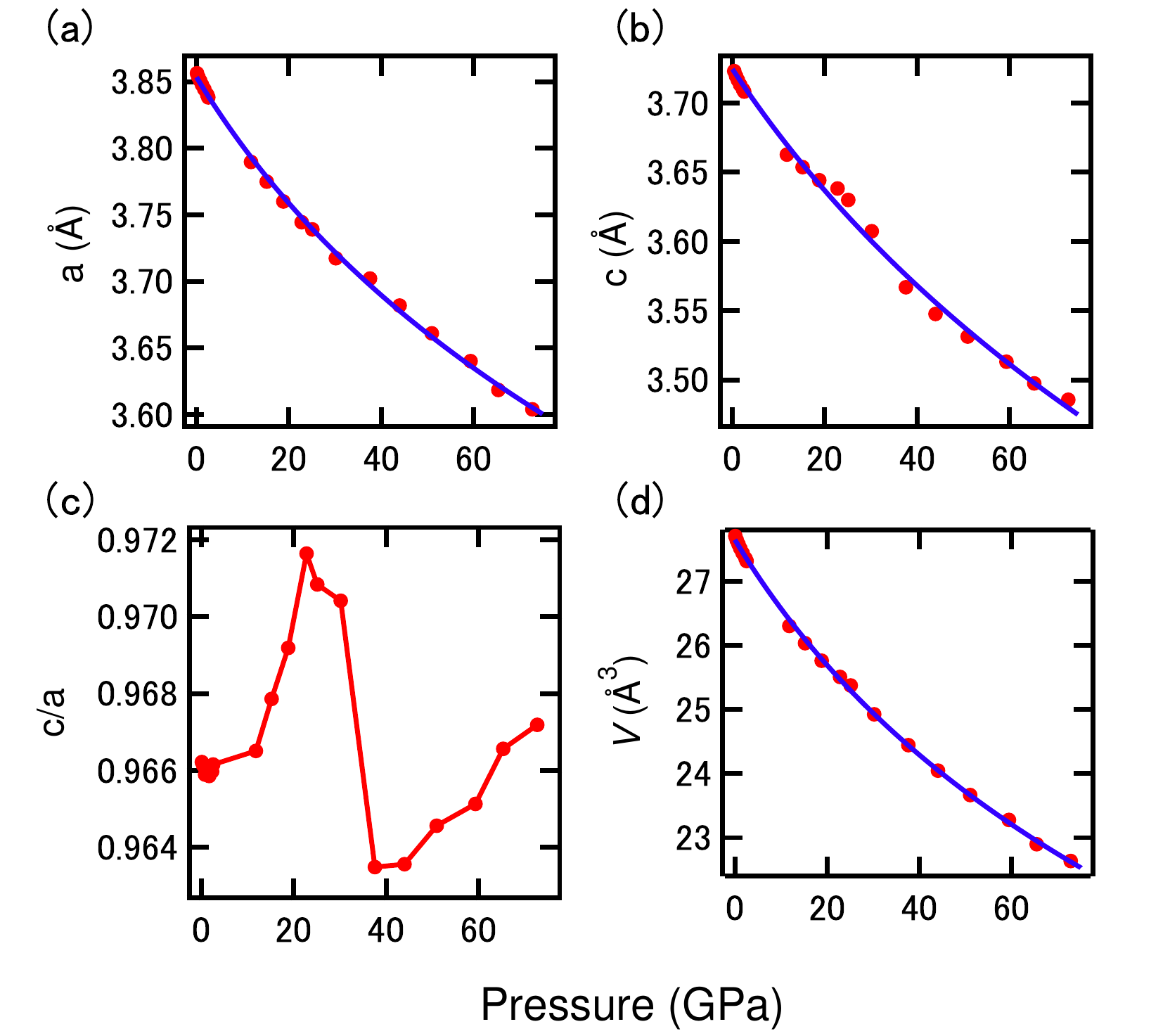}
\caption{Lattice constants ($a$ and $c$), 
the $c/a$ ratio, and the primitive-cell volume ($V$) 
obtained by powder XRD under high pressure. 
The solid lines are the fitted curves by the 
third-order BM-EOS.}
\label{fig5}
\end{figure}

We further attempted to observe the magnetic moment of Pt 
by high-pressure XMCD, as shown in Fig.~6. 
The inset shows the XAS and XMCD spectra 
at the Pt $L_3$ edge up to 20 GPa. 
One can see that the XMCD spectra are almost 
independent of pressure. The main panel shows the 
XMCD intensity at Pt $L_3$ and $L_2$ edges 
under high pressure. The intensity did not change 
by applying high pressure up to 20 GPa, indicating 
that the Pt magnetic moment is insensitive to pressure. 
The pressure-independent magnetic moment is 
consistent with the calculation in Fig.~3, 
in particular with SOC, 
where Pt magnetic moments are almost constant 
in the pressure regime of the present experiments. 
This strongly suggests that SOC should be taken 
into account in the calculation of magnetism. 
The sudden drop of the Pt magnetic moment, 
which we expect based on the first-principles 
calculations in Fig.~3, will be left 
for the future XMCD measurements 
under the pressure greater than 50 GPa. 

\begin{figure}
\centering
\includegraphics[width=\linewidth]{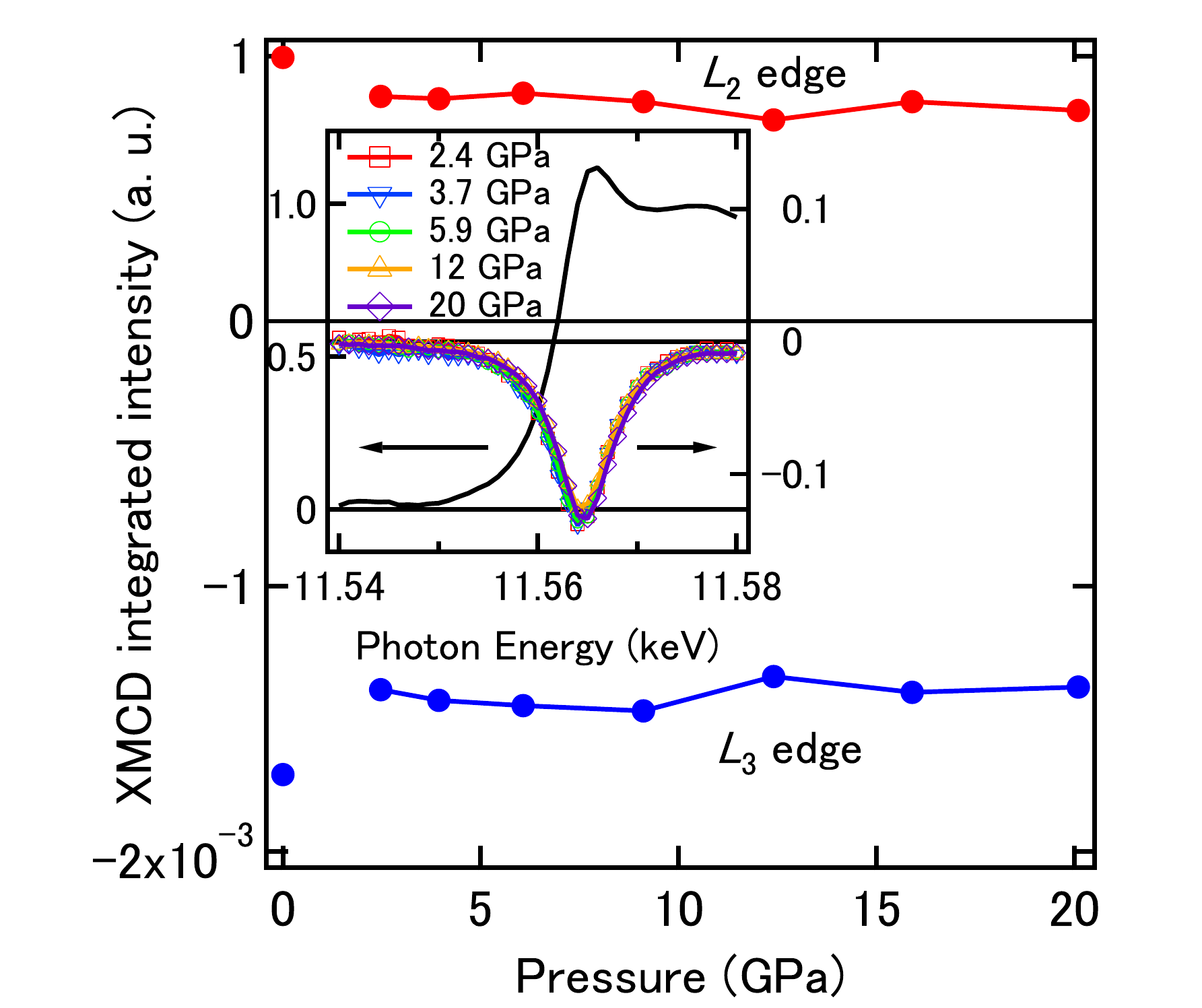}
\caption{The XMCD intensity at Pt $L_3$ and 
$L_2$ edges under high pressure. 
The inset shows the XAS (left axis) and 
XMCD (right axis) spectra at the Pt $L_3$ edge.}	
\label{fig6}
\end{figure}

In the present work, we studied the pressure dependence 
of the lattice constant and magnetism of $L1_0$-ordered FePt 
by first-principles calculations. 
Moreover, in the powder XRD and XMCD experiments, we 
observed most of the quantities that we 
numerically calculated. 
The magnetic anisotropy of $L1_0$-ordered FePt is considered 
to originate in SOC \cite{b3}, and thus it can be expected that 
the magnetic properties can be controlled in the future 
by changing the lattice constants under high pressure. 

\section{Summary}
We studied 
the relationship between the lattice constant and magnetism
of $L1_0$-ordered FePt under high pressure by 
a combination of first-principles calculations and 
x-ray measurements. 
The calculations showed 
the $c/a$ ratio anomaly at $\sim$ 20 GPa, 
which was observed by powder XRD under high pressure. 
We also measured the Pt magnetic moment by XMCD 
and found almost no pressure dependence, 
in agreement with the calculations. 
The XRD under higher pressure at $\ge$ 50 GPa 
is expected to reveal the decrease 
of the Pt magnetic moment. 
We successfully clarified the behavior of 
the lattice constants and magnetism in $L1_0$-ordered FePt 
under high pressure, which will lead to the creation of 
new magnetism in these materials. 

\section*{Acknowledgment}
We would like to thank S. Nakata 
for fruitful discussions. Measurements were performed 
with the approval of the Japan Synchrotron Radiation Research 
Institute (No. 2019B1623). This work was partially supported 
by JSPS KAKENHI to HW (JP19H05824) and to HF (JP19H02004).

\newpage
\begin{table*}
\renewcommand{\arraystretch}{1.3}
\caption{Lattice constants ($a$ and $c$), 
the $c/a$ ratio, and the primitive-cell volume ($V$) 
obtained by powder XRD under high pressure.}
\begin{center}
\begin{tabular*}{8.25cm}{|c|c|c|c|c|}
\hline 
Pressure (GPa) & $a$ (\AA) & $c$ (\AA) & $c/a$ & $V$ (\AA$^3$)\\ 
\hline 
0.1 & 3.8563(3) & 3.7261(5) & 0.9662(1) & 27.706(6)\\
0.4 & 3.8535(3) & 3.7230(5) & 0.9661(1)	& 27.642(6)\\
0.8	& 3.8507(6) & 3.719(1)  & 0.9658(3) & 27.57(1)\\
1.2	& 3.8476(8) & 3.717(2)	& 0.9661(6)	& 27.51(2)\\
1.7 & 3.844(1)  & 3.713(2)  & 0.9659(6) & 27.43(2)\\
2.3 & 3.840(1)  & 3.710(3)  & 0.9661(8) & 27.35(3)\\
2.5 & 3.838(1)  & 3.708(2)  & 0.9661(6) & 27.31(2)\\
11.8 & 3.790(1) & 3.663(2)  & 0.9665(6) & 26.31(2)\\
15.2 & 3.775(1) & 3.654(2)  & 0.9679(6) & 26.04(2)\\
18.8 & 3.7600(8)& 3.644(2)  & 0.9691(6) & 25.76(2)\\
22.8 & 3.744(1) & 3.638(2)  & 0.9717(6) & 25.50(2)\\
25.1 & 3.739(1) & 3.630(2)  & 0.9708(6) & 25.37(2)\\
30.2 & 3.717(1) & 3.607(3)  & 0.9704(8) & 24.92(2)\\
37.6 & 3.702(2) & 3.567(3)  & 0.964(1) 	& 24.44(3)\\
44.0 & 3.682(2) & 3.548(3)  & 0.964(1) 	& 24.05(3)\\
51.0 & 3.661(1) & 3.531(3)  & 0.9645(9) & 23.66(2)\\
59.4 & 3.640(2) & 3.513(3)  & 0.965(1) 	& 23.27(3)\\
65.4 & 3.618(1) & 3.497(2)  & 0.9666(6) & 22.89(2)\\
72.8 & 3.604(1) & 3.486(2)  & 0.9673(6) & 22.64(2)\\
\hline
\end{tabular*}
\end{center}
\label{Table3}
\end{table*}
\begin{table*}
\caption{Birch-Murnaghan equation-of-state parameters 
$K_0$, $K^{\prime}$, and $V_0$ obtained 
by fitting calculated and experimental results.}
\begin{center}
\begin{tabular}{|c|c|c|c|} 
\hline
& $K_0$ (\si{GPa}) & $K^{\prime}$ & $V_0$ (\AA$^3$)  \\ 
\hline
Calc (with SOC) & 210(3)  & 4.5(1) & 27.83(2)\\
Calc (without SOC) & 210(4) & 4.5(2) & 28.04(2)\\
Exp & 231(7) & 4.5(3) & 27.64(3)\\
\hline
\end{tabular}
\end{center}
\label{Table1}
\end{table*}
\end{document}